\documentclass[onecolumn, prd, aps, tightenlines, preprintnumbers, showpacs, nofootinbib, superscriptaddress, notitlepage]{revtex4-1}

\pdfoutput=1

\usepackage{amsmath}
\usepackage{color}
\usepackage{graphicx}
\usepackage{url}
\usepackage{epsfig}
\usepackage[T1]{fontenc}
\usepackage{multirow}

\usepackage{grffile}
\usepackage{amsmath,amsthm,amssymb,bm,amsfonts}
\usepackage{slashed}
\usepackage[utf8]{inputenc}
\usepackage{hyperref}

\usepackage{color}
\usepackage[normalem]{ulem}

\def\s0{\sigma_0}
\def\beq{\begin{equation}}
\def\eeq{\end{equation}}
\def\bear{\begin{eqnarray}}
\def\enar{\end{eqnarray}}

\newcommand{\state}[4]{{^#1\hspace{-0.6mm}#2_{#3}^{[#4]}}}

\newcommand\lamTheta{ \lambda_{\theta} }
\newcommand\lamPhi{\lambda_{\phi}}
\newcommand\lamThPh{\lambda_{\theta\phi}}

\newcommand\gev{\mathrm{~GeV}}
\newcommand\tev{\mathrm{~TeV}}

\newcommand{\jpsi}{{J/\psi}}

\newcommand\CScSa{\state{3}{S}{1}{1}}

\newcommand\COaSz{\state{1}{S}{0}{8}}

\newcommand\COcSa{\state{3}{S}{1}{8}}
\newcommand\COcPz{\state{3}{P}{0}{8}}

\newcommand\mo{{\mathcal O}}

\newcommand{\LDME}[2]{\langle\mo^{#1}(#2)\rangle}
\newcommand\mops{\LDME{\jpsi}{\CScSa}}
\newcommand\mopa{\LDME{\jpsi}{\COaSz}}
\newcommand\mopb{\LDME{\jpsi}{\COcSa}}
\newcommand\mopc{\LDME{\jpsi}{\COcPz}}

\newcommand{\vt}[1]{{{\boldsymbol #1}_\perp}}
\newcommand{\vtn}[2]{{{\boldsymbol #1}_{#2\perp}}}

\newcommand{\vp}{{\vt{p}}}

\newcommand{\vk}{{\vt{k}}}
\newcommand{\vka}{{\vtn{k}{1}}}

\newcommand{\vtp}[1]{{{\boldsymbol #1}'_\perp}}

\newcommand{\vkp}{{\vtp{k}}}

\begin{document}

\title{$\jpsi$ polarization in high multiplicity $pp$ and $pA$ collisions: CGC+NRQCD approach}

\author{Tomasz Stebel}
\email{tomasz.stebel@uj.edu.pl}
\affiliation{Institute of Theoretical Physics, Jagiellonian University, S.\L{}ojasiewicza 11, 30-348 Krak\'ow, Poland}

\author{Kazuhiro Watanabe}
\email{watanabe@jlab.org}
\affiliation{Theory Center, Jefferson  Laboratory, Newport News, Virginia 23606, USA}
\affiliation{SUBATECH UMR 6457 (IMT Atlantique, Universit\'e de Nantes, IN2P3/CNRS), 4 rue Alfred Kastler, 44307 Nantes, France}

\begin{abstract}

Quarkonium production mechanism in high multiplicity small collision systems has recently been pursued in the color-glass-condensate (CGC) effective theory combined with non-relativistic QCD (NRQCD) factorization, allowing to study initial state interactions. Quarkonium polarization, potentially measured in future experiments, would help elucidate the quarkonium production mechanism at high multiplicities. In this paper, we provide predictions on $\jpsi$ polarization parameters in high multiplicity proton-proton ($pp$) and proton-nucleus ($pA$) collisions within the CGC+NRQCD framework. Theoretical predictions are given for $\jpsi$ rapidity $2.5< y_\jpsi <4$, charged-particle multiplicity pseudorapidity $|\eta_{ch} | <1$ and energies $\sqrt{S}=13\tev$ for $pp$, $\sqrt{S}=8.16\tev$ for $pA$ collisions. Considering two leptonic frame choices (Collins - Soper and helicity) we find a weak polarization of $\jpsi$ that additionally decreases with growing event activities. No significant system size dependence between $pp$ and $pA$ collisions is obtained -- this could be a new constraint to initial state interactions in small collision systems.

\end{abstract}

\keywords{Quantum Chromodynamics}
\maketitle


\section{Introduction}

High multiplicity events in small collision systems ($pp$ and $pA$ collisions) at Relativistic Heavy Ion Collider (RHIC) and Large Hadron Collider (LHC) have received much attention in recent years. Many theoretical analyses have been performed to clarify whether a hot QCD medium can be present in such small collision systems; however, no clear consensus on this matter has been achieved thus far\,\cite{Dusling:2015gta,Schlichting:2016sqo,Nagle:2018nvi}.

Heavy quarkonium production in $pp$ and $pA$ collisions gives valuable opportunities to study the QCD dynamics in vacuum and medium\,\cite{Andronic:2015wma}. Experimental measurements about event activity dependence of quarkonium production at RHIC and the LHC are possible with good precision through dilepton decays\,\cite{Abelev:2012rz,Chatrchyan:2013nza,Adam:2015ota,Adam:2018jmp,Acharya:2020giw,Acharya:2020pit}.  As heavy quark pair is produced at an early stage in $pp$ and $pA$ collisions, quarkonium production mechanism can be modified in high multiplicity events compared to that in low multiplicity events due to either initial state effects or final state effects or both.

Heavy quarkonium production mechanism has been an exciting challenge to the QCD with multi-scales\,\cite{Brambilla:2010cs,Lansberg:2019adr}. Thanks to much efforts made over the last decade, hadronic quarkonium production at high transverse momentum $p_\perp$ can be described quantitatively in the NRQCD factorization\,\cite{Bodwin:1994jh}. Because of the large scale of $p_\perp$, the factorization between heavy quark pair production at a short distance and its transmutation into quarkonium at a long distance can be justified. Nevertheless, many issues remain to be resolved due to a lack of understanding of the long-distance matrix elements (LDMEs). Theoretical progress in the NRQCD at next-to-leading order (NLO) level\,\cite{Butenschoen:2011yh, Butenschoen:2012px,Chao:2012iv,Gong:2012ug,Shao:2014yta,Bodwin:2014gia,Bodwin:2015iua} has been confronted with inconsistency of the LDMEs extracted from data fitting in hadronic collisions\,\cite{Chung:2018lyq}. That is, relative weights between combinations of the LDMEs are unclear; see Ref.\,\cite{Zhang:2009ym} for discussion about the LDMEs fitted in $e^+e^-$ scatterings. Such an issue is also hidden when using the color evaporation model (CEM)\,\cite{Fritzsch:1977ay,Halzen:1977rs,Ma:2016exq,Cheung:2018tvq,Maciula:2018bex}, although this approach is not so well-motivated theoretically as the NRQCD.

The quarkonium production mechanism at low $p_\perp$ is another issue to be investigated closely. 
At unprecedented energy frontier with the LHC, highly occupied gluons with characteristic saturation scale $Q_s^2\gg\Lambda^2_\text{QCD}$ in hadron wave functions can be described by the CGC effective field theory\,\cite{Gribov:1984tu,Mueller:1985wy,Iancu:2003xm,Weigert:2005us,Gelis:2010nm,Kovchegov:2012mbw,Albacete:2014fwa,Blaizot:2016qgz}. 
The CGC framework allows to calculate amplitudes with resummed corrections from large logarithms $\alpha_s \ln(1/x)$, which could be predominant over leading logarithms in $Q^2$ when $p_\perp\lesssim Q_s\sim M_{\jpsi}\sim Q$\,\footnote{For $\Upsilon$ production at $p_\perp\lesssim Q_s\ll M_{\Upsilon}$, large logs in $Q^2$ are more significant\,\cite{Watanabe:2015yca}.}. Color field degrees of freedom in the hadron wave functions are separated into static color sources at large-$x$ and dynamical gauge fields at small-$x$. The requirement that cross-sections do not depend on this separation leads to the Jalilian-Marian-Iancu-McLerran-Weigert-Leonidov-Kovner (JIMWLK) equation\,\cite{JalilianMarian:1997gr,JalilianMarian:1997dw,Iancu:2000hn,Ferreiro:2001qy}, which is usually simplified using mean-field approximation into Balitsky-Kovchegov (BK) equation\,\cite{Balitsky:1995ub,Kovchegov:1999yj}. The $x$-evolution of the transverse momentum dependent gluon distribution in hadrons/nuclei obeys the nonlinear BK-JIMWLK evolution. It is worth noting that the CGC framework can recover the $k_T$-factorization\,\cite{Gelis:2003vh}, transverse momentum dependent factorization\,\cite{Altinoluk:2019fui,Fujii:2020bkl} and collinear factorization.

Within the CGC framework, the cross-section for heavy quark pair production in a dilute-dense system (i.e. $pA$ and forward $pp$) was computed in Refs.\,\cite{Blaizot:2004wv,Fujii:2006ab,Fujii:2013gxa}, and short distance coefficients (SDCs) in the CGC framework combined with the NRQCD factorization were calculated in Refs.\,\cite{Kang:2013hta,Qiu:2013qka}. Thus far, the CGC+NRQCD framework has been successful in describing the $p_\perp$ spectra of $\jpsi$ in minimum bias $pp$ and $pA$ collisions\,\cite{Ma:2014mri,Ma:2015sia} and polarization observables\,\cite{Ma:2018qvc}; in those analyses, a specific set of the LDMEs was applied - they had been obtained in Ref.\,\cite{Chao:2012iv}.

Furthermore, the event activity dependence of $J/\psi$ production yield was explored in the CGC framework in Ref.\,\cite{Ma:2018bax}. 
In the CGC picture, one can attribute high multiplicity events to rare parton configurations in hadrons and nuclei, participating in scattering processes. Such rare parton configurations can be set up by changing the $Q_s^2$ at large $x$\,\cite{Dusling:2013qoz,Dusling:2015rja,Schenke:2016lrs}, giving stronger multiple-scattering effects to the heavy quark pair. As to other approaches, Refs.\,\cite{Siddikov:2019xvf,Kopeliovich:2019phc} discussed the saturation effect on quarkonium production, while Ref.\,\cite{Gotsman:2020ubn} studied quarkonium production in a dilute system where the saturation effect is less important. Percolation approach\,\cite{Ferreiro:2012fb} shares the concept with the saturation approach. In addition, {\sc PYTHIA}8 event generator has been used to study multi-parton interactions\,\cite{Weber:2018ddv}, which are partially included in the CGC, and another event generator {\sc EPOS}3 can also examine the saturation effect\,\cite{Werner:2013tya}.  Of particular importance is that recent LHC data\,\cite{Acharya:2020pit} have shown that the initial state effects without final state flow effect can describe the event activity dependence of $\jpsi$ yield quantitatively.

The polarization of the $\jpsi$ meson can be analyzed using its decay into lepton's pair. Three polarization parameters $\lamTheta$, $\lamPhi$, and $\lamThPh$ can be extracted from data by measuring an angular distribution of the leptons. Measurements of the polarization parameters at the Tevatron\,\cite{Affolder:2000nn,Abulencia:2007us} and later at RHIC\,\cite{Adare:2009js,Adamczyk:2013vjy,Adam:2020dcj} and the LHC\,\cite{Abelev:2011md,Chatrchyan:2013cla,Aaij:2013nlm,Acharya:2018uww} showed that $\jpsi$ is produced with almost no polarization from low to high $p_\perp$ in hadronic collisions. On the other hand, the first theoretical predictions based on collinear factorized color singlet production channel at leading order (LO) suggested substantially nonzero polarization at high $p_\perp$\,\cite{Braaten:1999qk}. The inclusion of the octet states in the NRQCD expansion with SDCs calculated using NLO pQCD\,\cite{Butenschoen:2012px,Chao:2012iv,Gong:2012ug,Shao:2014yta} reduced this discrepancy (known as ``$J/\psi$ polarization puzzle''), but no full agreement with polarization measurements has been reached yet. 
The polarization of hadronic inclusive $\jpsi$ production was also studied within the CGC+NRQCD framework\,\cite{Ma:2018qvc}; good overall description of minimum bias data on $\jpsi$ polarization in $pp$ collisions at forward rapidities from LHCb and ALICE\,\cite{Aaij:2013nlm,Abelev:2011md,Acharya:2018uww} was obtained. Moreover, the STAR experiment has recently reported\,\cite{Adam:2020dcj} nice agreement of the CGC+NRQCD predictions with data in minimum bias $pp$ collisions.

In this paper, we study the polarization of $\jpsi$ in high multiplicity small collision systems, which should provide us with useful information on the fundamental quarkonium production mechanism. For this purpose, we use the CGC+NRQCD framework following an approach from Ref.\,\cite{Ma:2018bax} and clarifying how much the initial state saturation effect could impact the polarization of $\jpsi$ as charged multiplicity increases. Up to now, there are no experimental data for $\jpsi$'s polarization as a function of charged particle multiplicity, so this paper aims to provide predictions for the future LHC experiments including High-Luminosity LHC\,\cite{Chapon:2020heu} with the kinematical conditions used in similar measurements for unpolarized $\jpsi$ production.

The paper is organized as follows. In section\,\ref{sect_jpsi_CGC+NRQCD}, we briefly present calculations of the polarization parameters in the CGC+NRQCD framework. In section\,\ref{sect:Incl_had_prod}, we discuss the charged hadron production within the CGC framework. Section\,\ref{sect_results} demonstrates numerical results about the polarization of $\jpsi$ in small collision systems. We then conclude in section\,\ref{sect_summary}.

\section{Quarkonium production in the CGC+NRQCD framework}
\label{sect_jpsi_CGC+NRQCD}

We begin with reviewing steps toward calculating quarkonium production (particularly $\jpsi$) in a dilute-dense system in the CGC+NRQCD approach. 

The spin density matrix elements, depending on the quarkonium transverse momentum $\vp$ and rapidity $y$, are calculated in the NRQCD factorization framework\,\cite{Bodwin:1994jh}:
\beq
\frac{d {\sigma}^{\jpsi}_{ij}}{d^2\vp dy} =\sum_\kappa \frac{d \hat{\sigma}_{ij}^\kappa}{d^2\vp d
y} \, \langle\mathcal O_\kappa\rangle\,,
\label{NRQCD_expansion}
\eeq
where the matrix indices $i$ and $j$ represent helicity of the $\jpsi$ in the amplitude and conjugated amplitude, respectively. In the above expression $d \hat{\sigma}^\kappa/d^2\vp dy$ denotes SDCs for heavy quark pair production in a specific spin-color state represented by $\kappa$. $\langle\mathcal O_\kappa\rangle$ are LDMEs that describe the nonperturbative transition between a $c\bar c$ pair and $\jpsi$ meson. As nonperturbative quantities they need to be determined by data fitting. Meanwhile, LDMEs should be process-independent in the NRQCD factorization.
For the $\jpsi$ production, most dominant intermediate states in the quark velocity expansion are 
\beq
^3S_1^{[1]}, \ ^1S_0^{[8]}, \ ^3S_1^{[8]}, \ ^3P_J^{[8]} \textrm{ with } J=0,1,2\,,
\label{quantum_states_for_J/psi}
\eeq
where standard spectroscopic notation is used for state $\kappa$: $\state{{2S+1}}{L}{J}{C}$ with $[C]$ denoting a color state (singlet $[1]$ or octet $[8]$).

In the CGC framework, the SDC for a color singlet $c\bar{c}$ production in a dilute-dense system (e.g. $pA$) is given by\,\cite{Ma:2018qvc}:
\bear\label{eq:dsktCS}
\frac{d \hat{\sigma}_{ij}^\kappa}{d^2\vp dy}&\overset{\text{CS}} =&\frac{N_c (\pi R_p^2)(\pi R_A^2)}{4(2\pi)^{9}
(N_c^2-1)} \underset{\vka,\vk,\vkp}{\int}
\tilde{\mathcal{N}}_{x_1}(\vka) \mathcal{N}_{x_2}(\vk)\mathcal{N}_{x_2}(\vkp) 
\nonumber \\
&\times& 
\mathcal{N}_{x_2}(\vp-\vka-\vk-\vkp)\,
{\cal G}^\kappa_{ij}\left(x_1,x_2, p,\vka,\vk,\vkp \right),
\enar
and for a color octet state by
\bear\label{eq:dsktCO}
\frac{d \hat{\sigma}_{ij}^\kappa}{d^2\vp d
y}&\overset{\text{CO}}=&\frac{N_c(\pi R_p^2)(\pi R_A^2)}{4(2\pi)^{7}
(N_c^2-1)} \underset{\vka,\vk}{\int}
\tilde{\mathcal{N}}_{x_1}(\vka) \mathcal{N}_{x_2}(\vk) \nonumber \\
&\times&\mathcal{N}_{x_2}(\vp-\vka-\vk)
\,\Gamma^{\kappa}_{ij}\left(x_1,x_2, p,\vka,\vk\right),
\enar
with $\int_{\vk}=\int d^2\vk$. Matrices $\Gamma^{\kappa}_{ij}$ and $ {\cal G}^\kappa_{ij}$ describe couplings of the Wilson lines to the $c\bar{c}$ pair. For the details of calculations and explicit expressions, see Appendix B of Ref.\,\cite{Ma:2018qvc}. Their traces were calculated earlier in Refs.\,\cite{Kang:2013hta,Ma:2014mri}. In expressions above $\pi R_{p(A)}^2$ denotes the effective transverse area of the projectile proton (target nucleus)\,\cite{Ma:2014mri}. In this paper, as we shall focus on ratios of spin density matrix elements, the overall normalization factors are not considered below. Forward scattering amplitudes $\mathcal{N}_{x}$ and $\tilde{\mathcal{N}}_{x}$ correspond to the Fourier transform of the dipole correlator of light-like Wilson lines in the fundamental and adjoint representation respectively at $x_{1,2}=\sqrt{(2m_c)^2+p_{\perp}^2}e^{\pm y}/\sqrt{S}$ (see section \ref{sect:Incl_had_prod} for more details).

For the LDMEs we will apply values obtained in Ref.\,\cite{Chao:2012iv} by fitting NLO collinear factorized pQCD+NRQCD results to the Tevatron high $p_\perp$ prompt $\jpsi$ data. This set of LDMEs was also used in the previous CGC+NRQCD studies, where a good description of data was obtained for the $p_\perp$ spectra of $\jpsi$ in $pp$ and $pA$ collisions\,\cite{Ma:2014mri,Ma:2015sia} and $\jpsi$ polarization in $pp$ collisions\,\cite{Ma:2018qvc}. The color singlet LDME is estimated using the value of the quarkonium wave-function at the origin in a potential model\,\cite{Eichten:1995ch}: $\mops=1.16/(2N_c) \gev^3$. The color octet LDMEs have the following values (with uncertainties): $\mopa=0.089\pm0.0098 \gev^3$, $\mopb=0.0030\pm0.0012 \gev^3$, and $\mopc/m_c^2=0.0056\pm0.0021\gev^3$, where $m_c$ is the charm quark mass for which we take the value $1.5\gev$. One practical reason for choosing this set of the LDMEs is that the numerical values for each of the color octet states are positive, so our computations are not conflicted with cancellations among the color octet contributions. 

From the spin density matrix elements\,\eqref{NRQCD_expansion} one obtains the polarization parameters:
\beq
\lamTheta=\frac{d\sigma^{\jpsi}_{11}-d\sigma^{\jpsi}_{00}}{d\sigma^{\jpsi}_{11}+d\sigma^{\jpsi}_{00}}\,, \hspace{1cm}
\lamPhi=\frac{d\sigma^{\jpsi}_{1,-1}}{d\sigma^{\jpsi}_{11}+d\sigma^{\jpsi}_{00}}\,, \hspace{1cm}
\lamThPh=\frac{\sqrt{2}\; \rm{Re}(d\sigma^{\jpsi}_{10}) }{d\sigma^{\jpsi}_{11}+d\sigma^{\jpsi}_{00}}\, .
\label{lam_Defin}
\eeq
Those three parameters can be measured experimentally by considering the leptonic decay of $\jpsi$ in its rest frame. By $\Omega=(\theta,\phi)$ we denote the solid angle of the positive lepton w.r.t. arbitrary chosen vectors $X,Y,Z$. If $Y$ is chosen to be perpendicular to the hadronic plane, the angular distribution of the measured positive lepton is given by \cite{Noman:1978eh,Lam:1978pu}:  
\beq
\label{ang_distribution_in_Jpsi_frame}
\frac{d\sigma^{\jpsi(\rightarrow l^+l^- )}}{d\Omega} \propto 1+\lamTheta\cos^2{\theta}+\lamPhi \sin^2\theta\cos2\phi+\lamThPh \sin 2\theta \cos\phi.
\eeq
The angular distribution \eqref{ang_distribution_in_Jpsi_frame} depends on the orientation of vectors $X,Z$ w.r.t. hadrons' momenta. There are several choices of frames used in literature. We will use the most popular two: the Collins--Soper\,\cite{Collins:1977iv} frame and the recoil (also called helicity) frame\,\cite{Jacob:1959at}. For the explicit definition of these frames see for example Refs.\,\cite{Ma:2018qvc,Beneke:1998re}.

\section{Inclusive hadron production in the CGC framework}
\label{sect:Incl_had_prod}

This section summarizes the calculation of charged hadron multiplicity in the CGC framework; for more details, see Ref.\,\cite{Ma:2018bax}.

The pseudo-rapidity dependence of charged hadron multiplicity can be obtained in our approach as:
\begin{align}
\frac{dN_{ch}}{d\eta}\sim \int\limits^1_{z_\text{min}}\frac{dz}{z^2} \int d^2\bm{p}_{h\perp}\,  D_h(z) J({y_h\rightarrow\eta})\frac{d\sigma_{g}}{d^2\bm{p}_{g\perp} dy_g},
\label{eq:Nch-dy}
\end{align}
where we have skipped the overall normalization constant as we are interested only in ratios of this quantity.
$d\sigma_{g}/d^2\bm{p}_{g\perp} dy_g$ denotes the gluon production cross section calculated in the CGC framework. 
For the fragmentation function $D_h(z)$ we take the NLO parametrization of the Kniehl-Kramer-Potter fragmentation function at $\mu=2$\;GeV\,\cite{Kniehl:2000fe}. The hadrons carry $z$ fraction of the gluon's transverse momentum: $\bm{p}_{h\perp}=z \bm{p}_{g\perp}$. The Jacobian $J({y_h\rightarrow\eta})$ accounts for the transformation between rapidity $y_h$ and pseudo-rapidity $\eta$ of the hadron. In addition we assume that the rapidity of the hadron is the same as the rapidity of the parent gluon: $y_h =y_g$. For the typical mass of a light hadron we take $m_h=0.3\gev$.  The integration over $z$ in \eqref{eq:Nch-dy} is constrained from below by the kinematical conditions $x_{1,2}\leq1$, where $x_{1,2}=p_{g\perp}e^{\pm y_g}/\sqrt{S}$.

In the CGC framework, the cross section for the inclusive gluon production reads\,\cite{Kovchegov:2001sc,Blaizot:2004wu}:
\begin{align}
\frac{d\sigma_{g}}{d^2\bm{p}_{g\perp} dy}\sim  \underset{\vka }{\int} \, \frac{\boldsymbol k^2_{1\perp} (\vka-\bm{p}_{g\perp})^2}{\bm{p}_{g\perp}^2} \, \tilde{\mathcal{N}}_{x_1}(\vka) \,  \tilde{\mathcal{N}}_{x_2}(\vka-\bm{p}_{g\perp}) \theta\left(\bm{p}_{g\perp}^2- \boldsymbol k^2_{1\perp}  \right).
\label{eq:gluon-kt}
\end{align}
For the numerical purposes we restrict values of $p_{h\perp}$ in \eqref{eq:Nch-dy} to $0.1\gev \le  |\bm{p}_{h\perp}| \le 10\gev$. This cut has negligible effect on the results since the cross section decreases rapidly with $\bm{p}_{g\perp}$ and, in addition, we shall consider only ratios of $d N_{ch}/{d\eta}$.

The dipole correlators $\tilde {\mathcal{N}}_{x_1}(\vk)$ and $\mathcal{N}_{x_2}(\vk)$, which appear in \eqref{eq:dsktCS}, \eqref{eq:dsktCO}, and \eqref{eq:gluon-kt}, can be obtained by solving the BK equation. In this paper we use numerical solutions of the running coupling BK (rcBK) equation\,\cite{Balitsky:2006wa} in momentum space\,\cite{Albacete:2012xq}. Initial conditions at $x_0=0.01$ were chosen according to McLerran--Venugopalan (MV) model\,\cite{McLerran:1993ni,McLerran:1993ka} which in the position space reads
\begin{align}
D_{x_0}(\bm{r}_\perp)=
\exp\left[-\frac{\left(r_\perp^2Q_{s0}^2\right)^\gamma}{4}\ln\left(\frac{1}{r_\perp\Lambda}+e\right)\right],
\ \textrm{where} \ 
D_{x}(\bm{r}_\perp)= \underset{\vk }{\int} e^{-i \vk\cdot\bm{r}_\perp} {\mathcal{N}}_{x}(\vk)\,.
\label{eq:IC-rcBK}
\end{align}
This initial condition is parametrized by saturation scale $Q_{s0}^2$ in the proton/nuclei at $x_0 =0.01$. $Q_{s0}^2$ is proportional to $Q_0^2$ -- the saturation scale of the proton for minimum bias events. Values 
$\Lambda=0.241\gev$, $\gamma=1.119$ and $Q_{0}^2=0.168\gev^2$ were obtained in Refs.\,\cite{Albacete:2009fh,Albacete:2010sy,Albacete:2012rx} by fitting to HERA DIS data with $x\leq x_0$.

In the dense saturation regime at high energies, hadron multiplicity per unit of rapidity and transverse area in impact parameter space in the central rapidity region for symmetric $pp$ collisions can scale as follows \cite{Krasnitz:2000gz,SchaffnerBielich:2001qj}:
$dN_{ch}/d\eta \propto \langle k_\perp^2\rangle S_{\perp}/\alpha_s \sim Q_s^2S_{\perp}/\alpha_s$
with $S_{\perp}$ being an effective transverse area for impact parameter, $\langle k_\perp \rangle$ the average transverse momentum of produced hadrons. The above scaling is a consequence of Eq.(\ref{eq:gluon-kt}). High multiplicities can be achieved due to the hot spots where gluons are highly occupied, giving larger saturation scales; such rare parton configurations are less likely to happen compared to minimum bias events. It has been demonstrated numerically in \cite{Ma:2018bax} that increasing the initial saturation scales gives rise to high multiplicities. Therefore, we use the value of $Q_{s0}^2$ in \eqref{eq:IC-rcBK} as a parameter controlling charged multiplicity of a given event. We assume that for $pp$ collisions the initial saturation scales in both protons are the same: $Q_{s0,\rm{proton \, 1}}^2=Q_{s0,\rm{proton \, 2}}^2\equiv Q_{s0,\rm{proton}}^2$\,\footnote{Note that $Q_{s0}^2$ is the initial saturation scale, hence at $x_0=0.01$. The ``evolved'' saturation scales probed in the given event are determined by the rcBK equation and depend on the $x$ values. By increasing the initial saturation scale, we also increase the saturation scales at smaller $x$, which lead to higher event activities.}. It is equal to $Q_{0}^2$ for minimum bias events and $\xi Q^2_{0}, \ \xi>1$ for high multiplicity events. So we have:
\beq
\bigg\langle \frac{dN^{pp}_{ch}}{d\eta} \bigg\rangle  \equiv \frac{dN_{ch}}{d\eta} \bigg|_{Q_{s0,\rm{proton}}^2 = Q^2_{0}}, 
\label{min_bias_pp}
\eeq
for minimum bias events, where l.h.s. is calculated using Eq.\,\eqref{eq:Nch-dy} and
\beq
 \frac{dN^{pp}_{ch}}{d\eta}   \equiv \frac{dN_{ch}}{d\eta} \bigg|_{Q_{s0,\rm{proton}}^2 =\xi Q^2_{0}} ,
\label{high_mult_pp}
\eeq
for higher multiplicity events, with $\xi>1$. From definition, $\frac{dN^{pp}_{ch}}{d\eta}>\left< \frac{dN^{pp}_{ch}}{d\eta} \right>$.

For $pA$ collisions, the initial saturation scale in the nucleus, which is embedded in the MV initial condition, is expected to be larger than in the proton, $Q_{s0,\rm{nucleus}}^2 = \nu Q_{s0,\rm{proton}}^2$ with $\nu>1$. In this case the scaling of $dN_{ch}/d\eta$ is expected to be more complicated than symmetric $pp$ collisions because there can be the dependence on $\ln(Q_{s,{\rm proton}}/Q_{s,{\rm nucleus}})$\,\cite{Dumitru:2001ux}. Nevertheless, we drop such a explicit logarithmic correction for simplicity and instead tune the initial saturation scales by hand \cite{Ma:2015sia,Fujii:2015lld,Fujii:2017rqa,Ma:2018bax}. We then have for minimum bias $pA$ collisions: 
\beq
\bigg\langle \frac{dN^{pA}_{ch}}{d\eta} \bigg\rangle  \equiv \frac{dN_{ch}}{d\eta} \bigg|_{Q_{s0,\rm{proton}}^2 = Q^2_{0};  \ Q_{s0,\rm{nucleus}}^2 = \nu Q^2_{0}}.
\label{min_bias_pA}
\eeq

For higher multiplicity events, we assume that the initial saturation scales grow by the same factor $\xi$ in the proton and nucleus:
\beq
\frac{dN^{pA}_{ch}}{d\eta}  \equiv \frac{dN_{ch}}{d\eta} \bigg|_{Q_{s0,\rm{proton}}^2 =\xi Q^2_{0};  \ Q_{s0,\rm{nucleus}}^2 = \xi \nu \, Q^2_{0}}, 
\label{high_mult_pA}
\eeq
with $\xi>1$. The parameter $\nu$, being the ratio of initial saturation scales in the nucleus and proton, is not derived from first principles, i.e., QCD. Some previous studies\,\cite{Dusling:2009ni,Ma:2015sia,Fujii:2015lld,Fujii:2017rqa,Ma:2017rsu} suggest that it should be between 2 and 3 for a heavier target, i.e., Pb. We will calculate results both for $\nu = 2 $ and $\nu = 3$, and treat this factor as an additional source of uncertainty.

As we described above, we use the rcBK solution when $x<x_0=0.01$. For $x>0.01$, we employ an extrapolation of the solutions of the rcBK equation by requiring that the corresponding integrated gluon distribution matches that in the collinear factorization framework, see \cite{Ma:2014mri,Ma:2018bax} for more details.

In what follows, we will calculate the ratios between the yield in high-multiplicity $pp$ ($pA$) collisions and that in minimum bias events:
\beq
\frac{dN^{pX}_{ch}}{\langle dN^{pX}_{ch}\rangle} \equiv 
\int_{\eta}  \frac{dN^{pX}_{ch}}{d\eta}   
\Bigg/
 \int_{\eta}  \bigg\langle \frac{dN^{pX}_{ch}}{d\eta} \bigg\rangle ,
\label{Nch/<Nch>formula}
\eeq
where $X=p,\, A$ and multiplicities are defined in Eqs.\,\eqref{min_bias_pp}--\eqref{high_mult_pA}. Both multiplicities are integrated over the same pseudorapidity range.

The polarization parameter $\lamTheta$ \eqref{lam_Defin} in $pA$ collisions can be written as:
\beq
\lamTheta  =  
\left.\left( 
\frac{ 
\int_{y, \, \vp}   \frac{d\sigma^{\jpsi}_{11}}{d^2\vp dy} -\frac{d\sigma^{\jpsi}_{00}}{d^2\vp dy}     
     }
{ 
\int_{y, \, \vp}    \frac{d\sigma^{\jpsi}_{11}}{d^2\vp dy} +\frac{d\sigma^{\jpsi}_{00}}{d^2\vp dy} 
} 
 \right)
\right|_{Q_{s0,\rm{proton}}^2 =\xi Q^2_{0};  \ Q_{s0,\rm{nucleus}}^2 = \xi \nu \, Q^2_{0}}.
\label{lambdaTheta_Jpsi_pp}
\eeq
where all density matrix elements \eqref{NRQCD_expansion} are evaluated using the same initial saturation scales. Just like for the light hadrons multiplicity, the parameter $\xi$ `controls' the activity of the events. Polarization parameters $\lamPhi$ and $\lamThPh$ \eqref{lam_Defin}, are calculated similarly to \eqref{lambdaTheta_Jpsi_pp}. The calculation for $pp$ collisions is analogous, with $\nu=1$, as in Eq.\,\eqref{high_mult_pp}.

As we have shown above, the $\jpsi$ production cross-section in the CGC+NRQCD framework \eqref{NRQCD_expansion} does not include any medium effect but the saturation effect. In this paper, we assume that the bound state formation happens at a later stage without modifying the LDMEs, which could be acceptable only when considering $\jpsi$ production\,\cite{Ma:2017rsu}. Thus any modifications in the $\jpsi$'s polarization in $pA$ collisions and high multiplicity events can be attributed to the saturation effect at a short distance. In the following section, we shall clarify the impact of the saturation effect on the polarization of $\jpsi$.

\section{Results}
\label{sect_results}

\begin{figure*}
\begin{center}
\includegraphics[width=.47\textwidth]{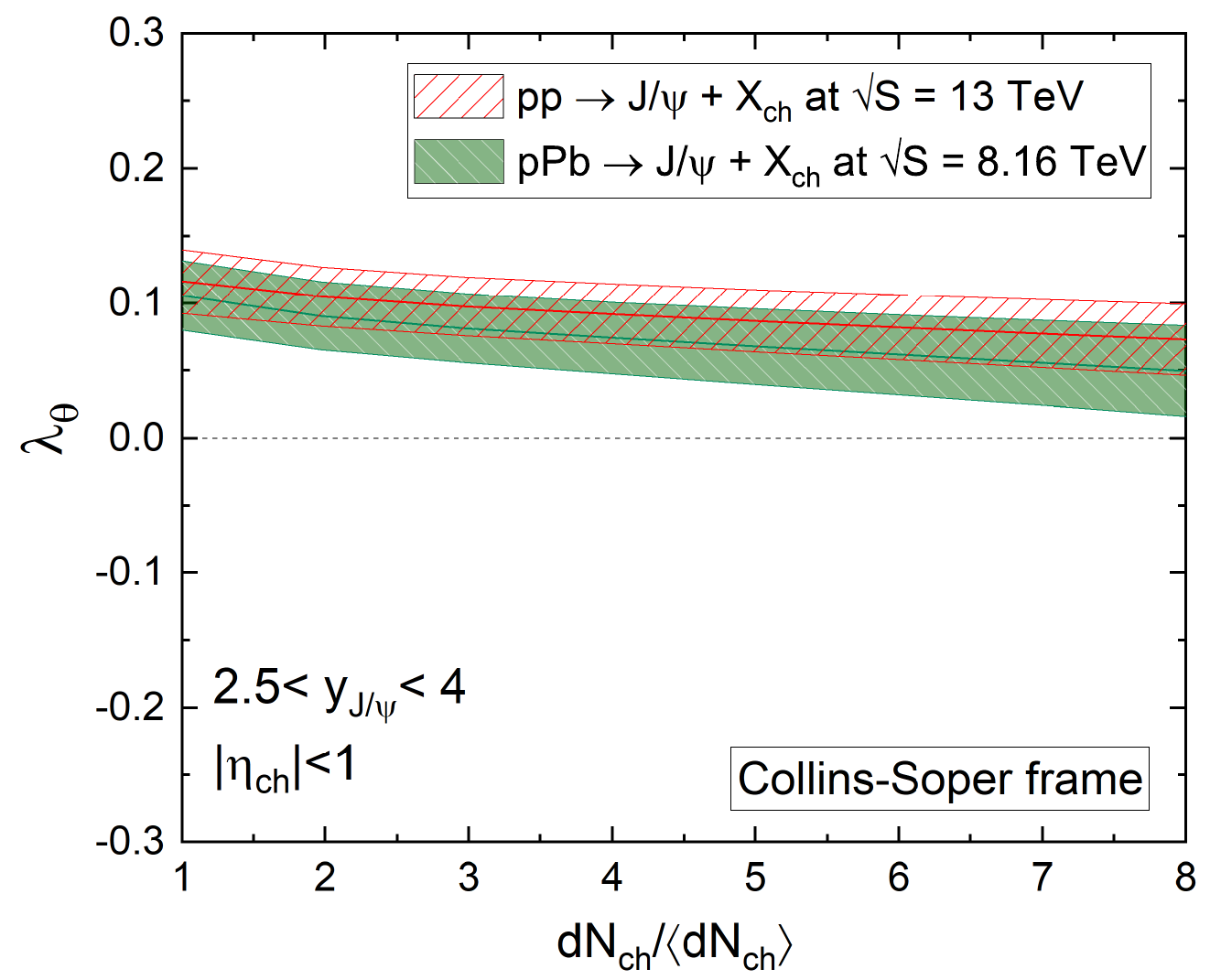}
\includegraphics[width=.47\textwidth]{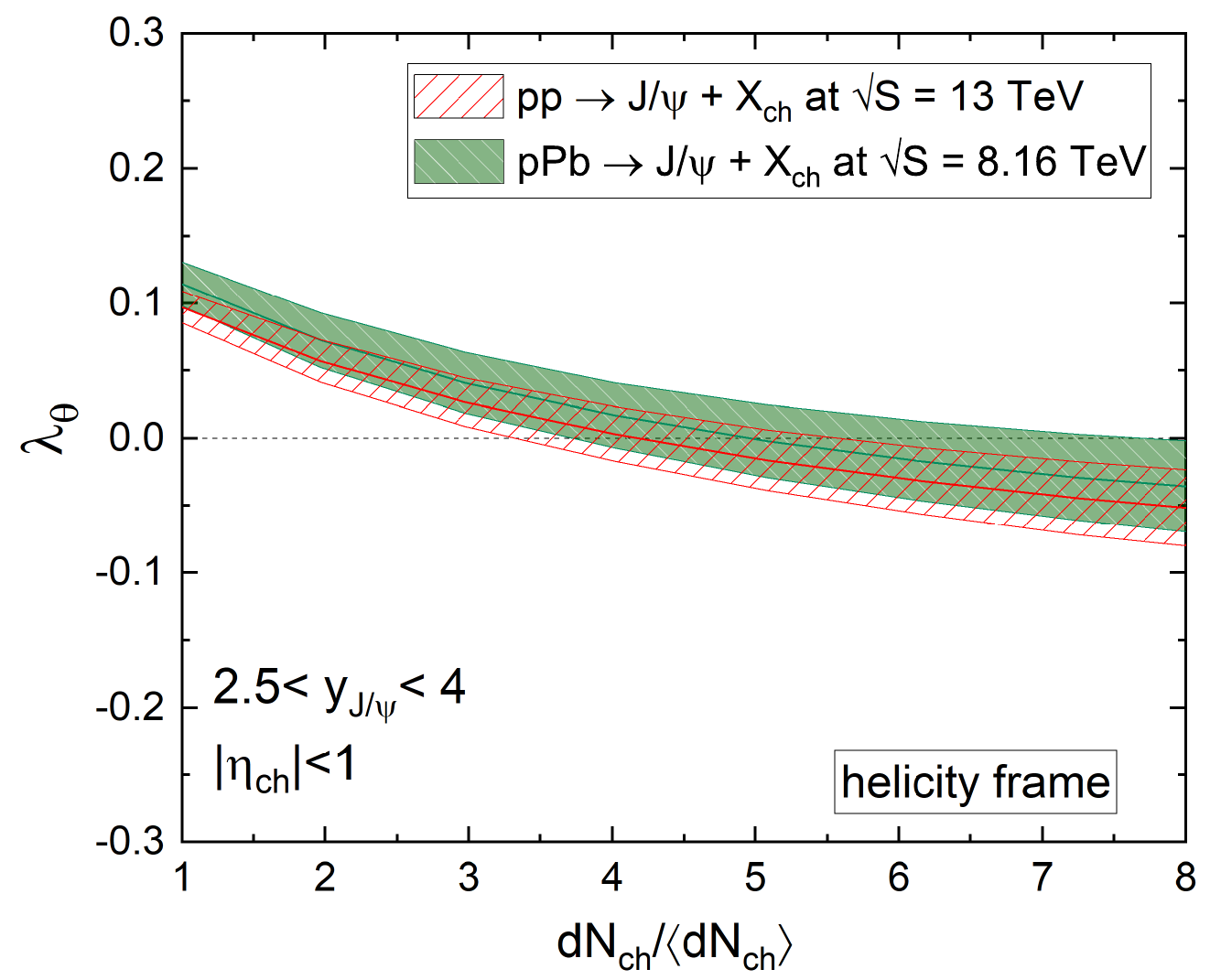}
\includegraphics[width=.47\textwidth]{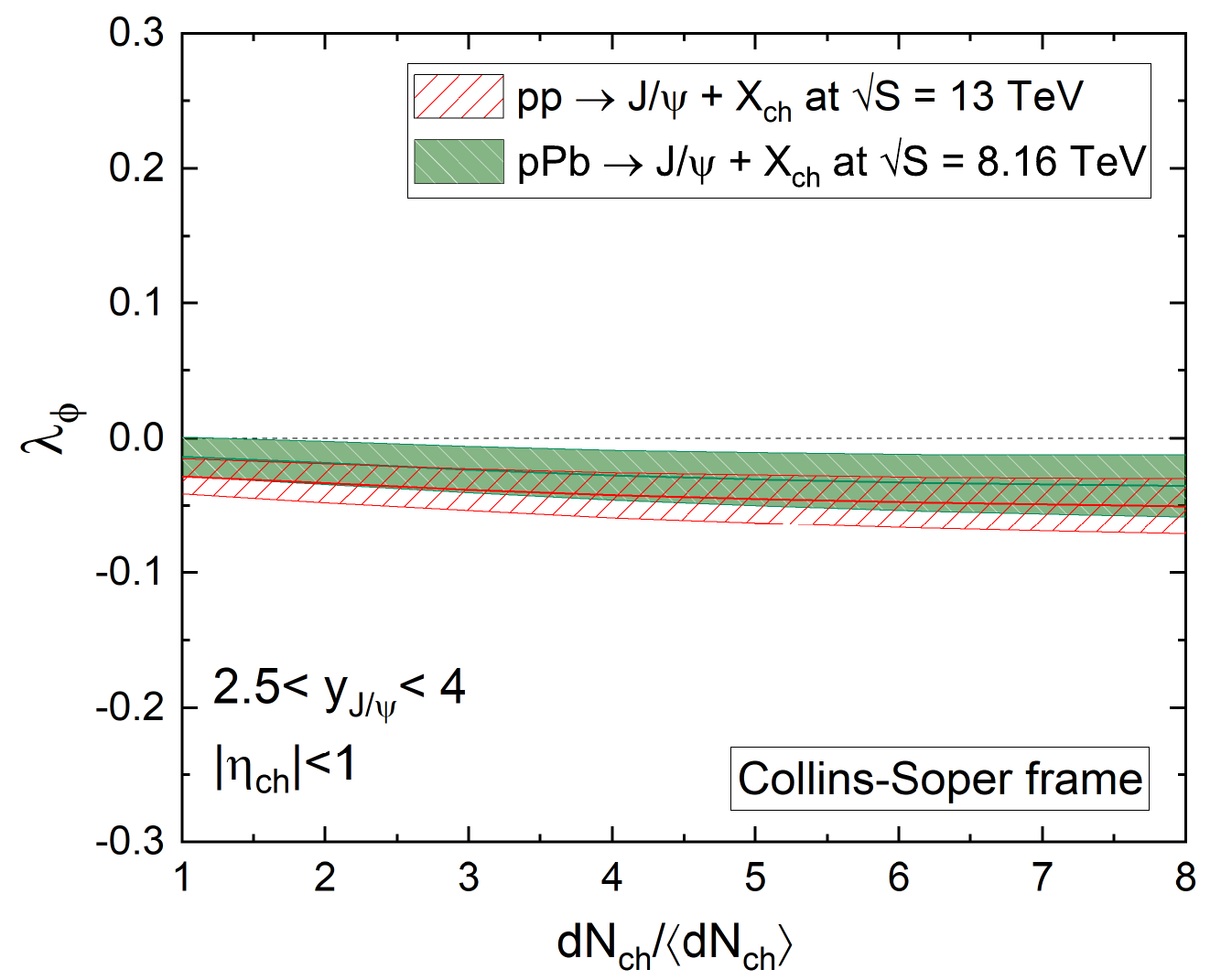}
\includegraphics[width=.47\textwidth]{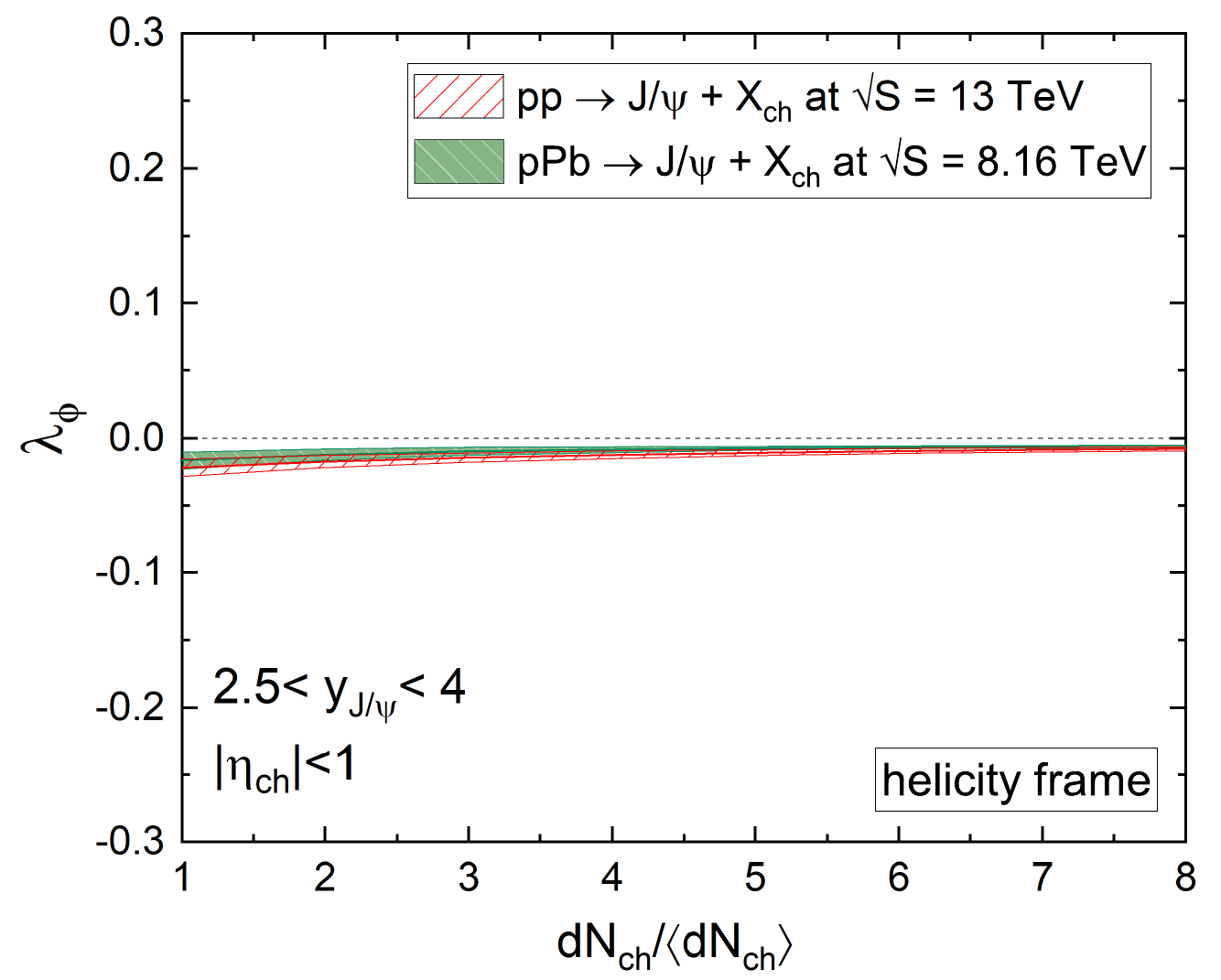}
\includegraphics[width=.47\textwidth]{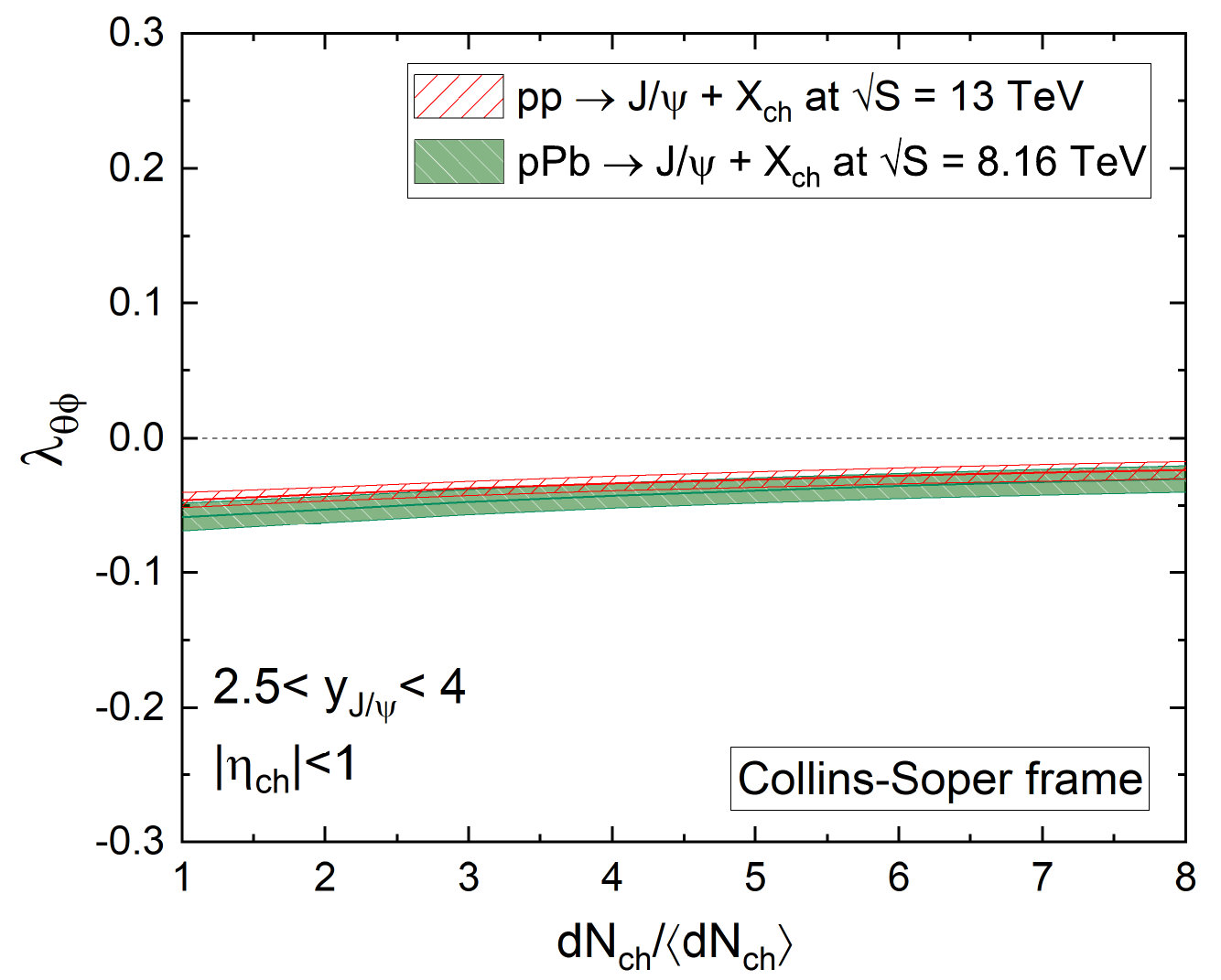}
\includegraphics[width=.47\textwidth]{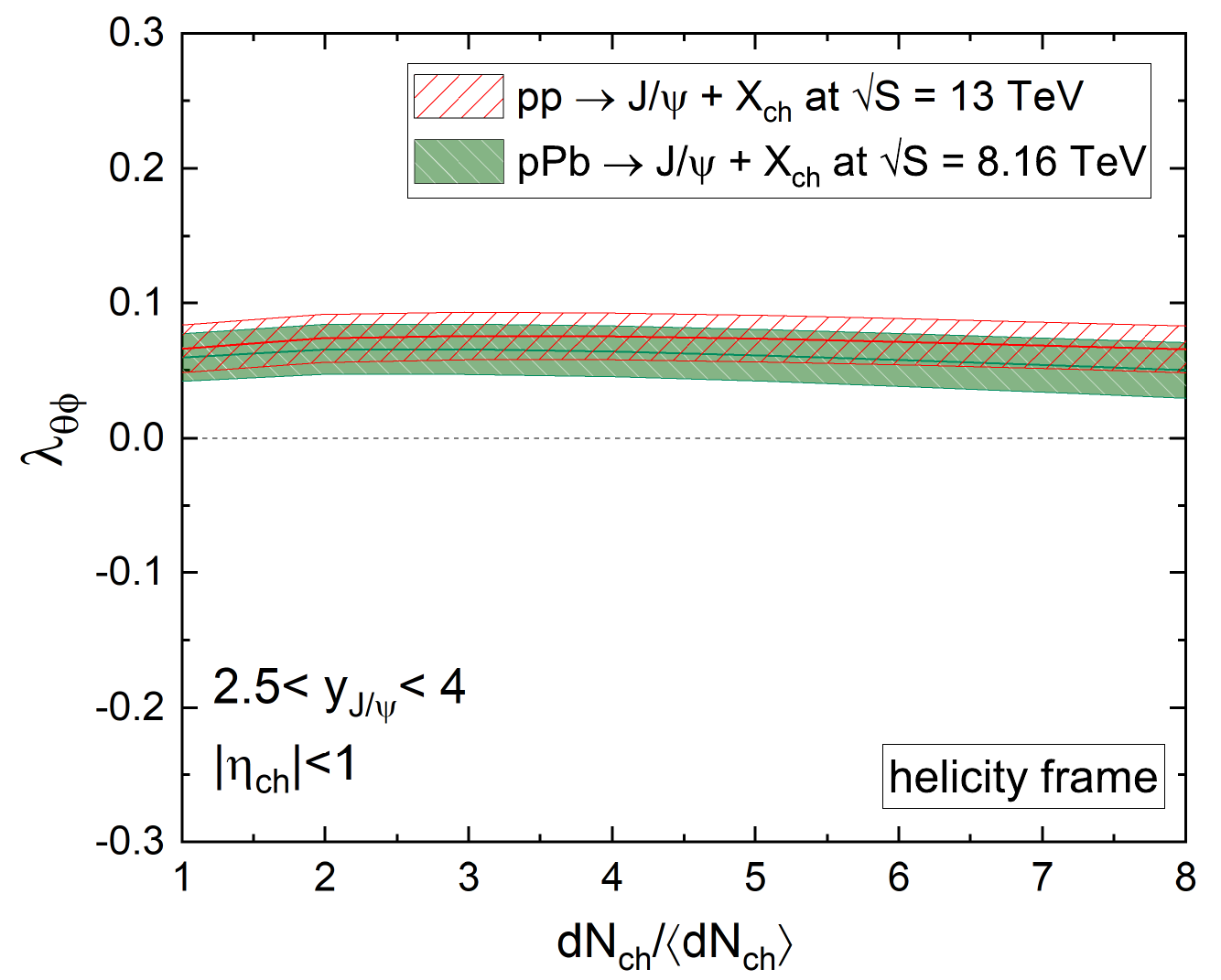}
\end{center}
\caption{Polarization parameters $\lamTheta$ (upper row), $\lamPhi$ (middle row) and  $\lamThPh$ (lower row) of forward $\jpsi$ production as a function of relative multiplicity in $pp$ collisions at 13 TeV (red hatched band) and $pA$ collisions at 8.16 TeV (green full band) collisions. Predictions are given for two helicity frames: Collins-Soper (left column) and helicity (right column). 
}
\label{lambda_plot}
\end{figure*}

Since experimental data for polarization parameters in high multiplicity events are not available at present, we will give our predictions with kinematical conditions probed in the related measurements. We choose energies from Run 2 of the LHC, $\sqrt{S}=13\tev$ for $pp$ collisions and $\sqrt{S}=8.16\tev$ for $pA$ collisions. We constrain the charged hadrons' pseudorapidity to the window $|\eta_{ch} | <1$, which was used in several ALICE measurements of $\jpsi$'s yield as a function of charged-particle multiplicity \cite{Abelev:2012rz,Acharya:2020giw,Acharya:2020pit}. The $\jpsi$ rapidity we restrict to the window $2.5< y_\jpsi <4$, where positive rapidity is defined by the proton-going direction (for $pA$ collisions). This setup provides the dilute projectile -- dense target scattering, where the CGC formalism is particularly effective. Moreover, this range of rapidity has already been used for the measurements of $\jpsi$ polarization\,\cite{Abelev:2011md,Acharya:2018uww}\footnote{In Ref.\,\cite{Ma:2018qvc} we have shown that in the CGC+NRQCD model, for the forward $\jpsi$ production, the polarization parameters are not sensitive to the rapidity.}.

In order to estimate uncertainties of our predictions, we use errors of the LDMEs (see section \ref{sect_jpsi_CGC+NRQCD}) - we vary each LDME by the value of its error and calculate maximal and minimal value of the polarization parameters obtaining a band of uncertainty. For the $pA$ collisions also the uncertainty from unknown ratio $\nu= Q_{s0,\rm{nucleus}}^2/Q_{s0,\rm{proton}}^2$ was included (see \eqref{high_mult_pA} and its discussion).

Our results for three polarization parameters are displayed in Figure\,\ref{lambda_plot}. We provide predictions for two leptonic frames: Collins--Soper frame (left column) and helicity frame (right column). For each polarization parameter, the predictions for $pp$ (red hatched band) and $pA$ (green filled band) collisions are shown. We have checked that results do not depend significantly on energy $\sqrt{S}$. This is consistent with the previous result that the correlation between the yield of $\jpsi$ and charged-hadron may be characterized by $Q_s^2$, so that the $\sqrt{S}$ dependence is weak\,\cite{Ma:2018bax}. The difference between $pp$ and $pA$ results in Figure\,\ref{lambda_plot} comes entirely from the different type of target. Values of all parameters do not exceed (in the module) 0.1, which means almost unpolarized production of $\jpsi$. The parameter $\lamTheta$ in both frames decreases with the event activity -- this behavior is particularly well seen in the helicity frame where $\lamTheta$ is equal to 0.1 at minimum bias events and decrease with the growing event activity to negative values. In the Collins--Soper frame, the $\lamTheta$ in $pA$ collisions is slightly smaller than in $pp$ collisions. In the helicity frame, the opposite is true. The $\lamPhi$ parameter is negative and very close to 0 in both frames. The $\lamThPh$ parameter is larger than $\lamPhi$, has a different sign in both frames, and very slightly decreases with the event activity. We note that the small polarization of produced $\jpsi$ in the CGC+NRQCD approach was also predicted for the inclusive events (see \cite{Ma:2018qvc}).

Using our approach, we predict a slight difference between the $\jpsi$ polarization in $pp$ collisions and that in $pA$ collisions. All the differences are within the uncertainty bands. This characteristic behavior can be traced back to the feature of our model. Firstly, the saturation effect participates in partonic scatterings at short distance, where a produced $c\bar c$ pair does not experience the hadronization effects. 
Secondly, the same values of the LDMEs were used both in $pp$ and $pA$ collisions -- this universality is suggested by the good description of the $p_\perp$ spectra of $\jpsi$ in $pp$ and $pA$ collisions with the same set of the LDMEs \cite{Ma:2014mri,Ma:2015sia}. Finally, in our model, there is no final state interaction of $\jpsi$. One should note that the similarity between $pp$ and $pA$ collisions is not present in data for $\jpsi$'s yield and mean $p_\perp$ at high multiplicity events. If we use the same set up shown in this paper, the CGC predictions for those observables differentiate between $pp$ and $pA$ collisions\,\cite{Ma:2018bax} because the $\jpsi$'s yield and $p_\perp$ are driven by combinations of the saturation scales, which are very different in the proton and nucleus. On the other hand, the polarization observables are not sensitive to those differences, as we have shown here. Therefore, we conclude that $\jpsi$'s polarization data will provide a complementary test of the CGC theory at the high multiplicity events.

\section{Summary}
\label{sect_summary}

In this paper, we have demonstrated the polarization parameters of $\jpsi$ as a function of charged-particle multiplicity in $pp$ and $pA$ collisions in the CGC framework. Our work is a continuation of the previous analyses, where the CGC framework was applied to high multiplicity events\,\cite{Ma:2018bax} and (joined with the NRQCD) to $\jpsi$ polarization studies\,\cite{Ma:2018qvc}. In this approach, the event activity is controlled by the value of the saturation scale encoded in the solution of the rcBK evolution equation.

We predict a very small polarization of the produced $\jpsi$ in minimum bias $pA$ collisions, which is almost consistent with the $\jpsi$ polarization in $pp$ collisions. According to our model, the polarization parameters decrease slowly as the event activity increases. We found that the $\jpsi$'s polarization has a weak dependence on the system size and $\sqrt{S}$, albeit the saturation scale in the target is different between $pp$ and $pA$ collisions. This is contrary to the other observables including $\jpsi$ yield and mean $p_\perp$\,\cite{Ma:2018bax}.

The polarization measurements of quarkonium in high multiplicity events provide an independent insight into the QCD dynamics behind high multiplicity events. Given the data which are currently available, such observable should be accessible experimentally at the LHC. Together with other measurements, like quarkonium's yield and mean $p_\perp$, polarization will help us understand the role of initial state interactions, hadronization, and final state interactions in quarkonium production processes.

\acknowledgments

We thank R.~Venugopalan and Y.-Q.~Ma for fruitful discussions and inspiration for this work. We also thank Anna Filipowska for useful comments on the manuscript. Support of the Polish National Science Center (NCN) Grant No.\,2019/32/C/ST2/00202 is kindly acknowledged. This work is supported by Jefferson Science Associates, LLC under U.S. DOE Contract No.\,DE-AC05-06OR23177.


\bibliography{biblio_Jpsi_CGC}

\end{document}